# Multiferroic properties of epitaxially stabilized hexagonal DyMnO$_3$ thin films

J.-H. Lee, P. Murugavel, D. Lee, and T. W. Noh[†]

*Research Center for Oxide Electronics, FPRD & Department of Physics and Astronomy,*

*Seoul National University, Seoul 151-747, Korea*

Y. Jo and M.-H. Jung

*Quantum Materials Research Team, Korea Basic Science Institute,*

*Daejeon 305-333, Korea*

K. H. Jang and J.-G. Park

*BK21 Physics Division, Department of Physics, Sungkyunkwan University, Suwon 440-746, Korea*

We fabricated epitaxial thin films of hexagonal DyMnO$_3$, which otherwise form in a bulk perovskite structure, via deposition on Pt(111)//Al$_2$O$_3$ (0001) and YSZ(111) substrates: each of which has in-plane hexagonal symmetry. The polarization hysteresis loop demonstrated the existence of ferroelectricity in our hexagonal DyMnO$_3$ films at least below 70 K. The observed 2.2 μC/cm$^2$ remnant polarization at 25 K corresponded to a polarization enhancement by a factor of 10 compared to that of the bulk orthorhombic DyMnO$_3$. Interestingly, this system showed an antiferroelectric-like feature in its hysteresis loop. Our hexagonal DyMnO$_3$ films showed an antiferromagnetic Néel temperature around 60 K and a spin reorientation transition around 40 K. We also found a clear hysteresis in the temperature dependence of the magnetization, which was measured after zero-field-cooling and field-cooling. This hysteresis may well have been of spin glass origin, which was likely to arise from the geometric frustration of antiferromagnetically-coupled Mn spins with an edge-sharing triangular lattice.

[†]E-mail: twnoh@snu.ac.kr

Research on multiferroic materials has been recently of significant interest due to their scientifically intriguing properties as well as potential applications.[1-9] Especially, lanthanide-based $R$MnO$_3$ ($R$= Tb, Dy,…, and Lu) compounds[3-9] have attracted a great deal of attention. These compounds form in two distinct crystal structures, depending on the ionic radius $r_R^{3+}$ of the rare-earth ions:[3] an orthorhombic structure (*Pbnm*) when $r_R^{3+} \geq r_{Dy}^{3+}$, and a hexagonal structure (*P6$_3$cm*) when $r_R^{3+} < r_{Dy}^{3+}$. In the orthorhombic $R$MnO$_3$ ($R$ = Tb, Dy, and Gd) compounds, the antiferromagnetic (AFM) transition temperature $T_N$ is reported to be around 40 K. Interestingly enough, their ferroelectricity was attributable to lattice modulation accompanied by the antiferromagnetic order.[4,5] Therefore they have a ferroelectric (FE) Curie temperature $T_C$ lower than $T_N$, and a small remnant polarization $P_r$ (< 0.2 µC/cm$^2$). However, in the hexagonal $R$MnO$_3$ ($R$ = Ho, Tm, Lu, and Y) compounds, $T_N$ values are around 70 K.[6-8] Their ferroelectricity is known to arise from a structural distortion[9] having $T_C$ much higher than 300 K and saturated polarization $P_s$ larger than 5.6 µC/cm$^2$.[6-8]

Note that the difference between the formation energies of the orthorhombic and the hexagonal $R$MnO$_3$ phases could be small. It might then be possible to grow a thin manganite film in a metastable phase, a phase that does not exist in the bulk form. If we can grow the film epitaxially to match the substrate's crystal symmetry, we can lower surface energy due to coherent film-substrate interface. For example, the hexagonal $R$MnO$_3$ (R=Y, Ho, Tm, and Lu) phases were reported to be stabilized as in metastable orthorhombic thin film forms artificially grown on top of substrates with an orthorhombic symmetry.[10] Similarly, it could be possible to stabilize orthorhombic $R$MnO$_3$ into a metastable hexagonal phase using a suitable substrate having a hexagonal symmetry.[11] As DyMnO$_3$ is an end member of the orthorhombic manganites, it would be an ideal candidate to engineer into hexagonal thin films and to search for newly emerging physical properties.

In this letter, we demonstrate that hexagonal DyMnO$_3$ films can be grown epitaxially on hexagonal in-plane symmetric Pt(111)//Al$_2$O$_3$(0001) and (111) oriented yttria-stabilized zirconia (YSZ) substrates. We found that the FE properties of these films are significantly enhanced compared to those of their orthorhombic bulk counterparts. In addition, the hexagonal DyMnO$_3$ films also reveal interesting physical phenomena, such as an antiferroelectric(AFE)-like signature above 70 K and spin-glass-like magnetic behavior.

We grew hexagonal DyMnO$_3$ thin films using the pulsed laser deposition technique. The DyMnO$_3$ target was ablated using a KrF excimer laser (248 nm, Lambda Physik) with a 1.5 J/cm$^2$ laser fluence and a 4 Hz repetition rate. During the deposition, the substrates were kept at 900 °C and the chamber oxygen partial pressure was maintained at 100 mTorr. After deposition, the thicknesses of the DyMnO$_3$ thin films were around 50 nm, as measured under a cross-sectional scanning electron microscope. A four-circle high-resolution X-ray diffractometer (XRD) was used to verify the epitaxial growth of the hexagonal DyMnO$_3$ films. The FE properties of the hexagonal DyMnO$_3$ films on Pt(111)//Al$_2$O$_3$(0001) were investigated with a low-temperature probe station (Desert Cryogenics) and T-F analyzer (aixACCT). The magnetic properties of the DyMnO$_3$ films on the YSZ(111) substrates were measured using a commercial SQUID magnetometer (Quantum Design).

From the structural studies, we confirmed that our hexagonal DyMnO$_3$ films were epitaxially grown with their c-axes normal to the substrate planes. Figures 1(a) and (b) show the XRD $\theta$-2$\theta$ scans of the DyMnO$_3$ films on Pt(111)//Al$_2$O$_3$(0001) and

YSZ(111) substrates, respectively. In the XRD data, we could only see the (0002) and (0004) reflections of the hexagonal phase along with the substrate peaks, indicating the formation of a pure hexagonal DyMnO$_3$ phase. The inset of Fig. 1(a) shows that the full-width-at-half-maximum in the rocking curve of the (0004) peak is about 0.63°, suggesting a reasonably good crystalline quality. Figures 1(c) and (d) show the XRD $\phi$-scans around the hexagonal DyMnO$_3$ (11$\bar{2}$2) peak for the films on the Pt(111)//Al$_2$O$_3$(0001) and YSZ(111) substrates, respectively. The presence of six distinct peaks with 60° separations indicates the epitaxial growth of the hexagonal films. From these XRD studies, it is clear that the hexagonal DyMnO$_3$ films were grown on Pt(111)//Al$_2$O$_3$(0001) and YSZ(111) substrates with epitaxial relationships of DyMnO$_3$[100]//Pt[11$\bar{2}$]//Al$_2$O$_3$[11$\bar{2}$0] and DyMnO$_3$[100]//YSZ[11$\bar{2}$], respectively.

To investigate the FE properties of the hexagonal DyMnO$_3$ phase, we measured the polarization-electric field (*P-E*) and dielectric constant-electric field ($\varepsilon$-*E*) hysteresis loops of the film grown on Pt(111)//Al$_2$O$_3$(0001). We used 100 μm diameter gold dots as top electrodes. The *P-E* loops measured at 25 and 140 K are shown in Figs. 2(a) and 2(b), respectively. The corresponding $\varepsilon$-*E* curves are shown in the insets of Fig. 2. With the maximum applied electric field of 2.4 MV/cm, $P_r$ and $P_s$ at 25 K were found to be about 2.2 and 9.8 μC/cm$^2$. Note that the reported $P_r$ value of bulk orthorhombic DyMnO$_3$ is about 0.2 μC/cm$^2$ at 5 K, which is at least one order of magnitude smaller than the values obtained in our hexagonally stabilized DyMnO$_3$ film.

The hexagonal DyMnO$_3$ films also showed AFE-like signatures in their *P-E* and $\varepsilon$-*E* loops above ~70 K. To provide a clearer picture, we evaluated d$^2$*P*/d*E*$^2$ curves from the upper polarization loop. As shown in the inset of Fig. 2(a), a single inflection point could be seen in the FE hysteresis curve (at 25 K). However, as shown in the inset of Fig. 2(b), three inflection points were observed in the $d^2P/dE^2$ curve for the 140 K data, corresponding to an AFE-like signature. Figure 3 shows the contour map for $P_r$ in the temperature *vs.* the maximum applied E-field plane during the *P-E* measurements. The boundary between FE and AFE-like phase was drawn by estimating the number of inflection points in the $d^2P/dE^2$ curve. Although the existence of the AFE phase has been predicted at high temperatures for some bulk hexagonal rare-earth manganites,[8,12] the origin of the AFE-like feature in our hexagonal $DyMnO_3$ films needs to be elucidated.

Since Pt electrode of the Pt(111)/Al$_2$O$_3$(0001) structure had rather high background signal in our magnetization measurements, it was difficult for us to reliably measure weak magnetic signals from the $DyMnO_3$ films. We therefore used a $DyMnO_3$ film grown on a YSZ(111) substrate without Pt electrode for magnetic studies shown below. Figure 4(a) shows the out-of-plane magnetization *vs.* temperature taken in a 100 Oe magnetic field under zero-field- cooled (ZFC) and field-cooled (FC) conditions. The open and solid circles correspond to the ZFC and FC data, respectively. Interestingly, both ZFC and FC magnetization data show two anomalies around 40 and 60 K before increasing rather sharply. We identified the transition around 60 K as the usual AFM Néel ordering of Mn moments with a 120º structure.[7,13] However, we believe that the second transition temperature, around 40 K, may have been due to a Mn spin reorientation as in bulk hexagonal $HoMnO_3$.[14]

It is noteworthy that a large difference existed between the ZFC and FC data, which, except for $ScMnO_3$, has not been observed in other hexagonal manganites.[15] We attributed this thermal hysteresis to spin-glass-like behavior, which might have originated from the strong geometrical frustration of antiferromagnetically-coupled Mn spins with an edge-sharing triangular lattice. As shown in Fig. 4(b), the in-plane magnetization also indicated similar anomalies. The strong increase in the magnetization at lower temperatures was most likely due to ordering of Dy moments. It is also interesting to note that the field dependence of the magnetization showed a clear hysteresis at 5 K, indicating that Dy might order ferromagnetically. This hysteresis in M(H) disappeared at higher temperatures.

In summary, we fabricated hexagonal $DyMnO_3$ thin films, whose bulk stable phase should be orthorhombic, using substrates with hexagonal in-plane atomic arrangements. Compared to the orthorhombic single crystal, the hexagonal $DyMnO_3$ thin films showed enhanced multiferroic properties. Their ferroelectric remnant polarization values have increased by more than one order of magnitude, along with a shift in ferroelectric transition temperatures, from 28 to ~70 K. The magnetic properties of the hexagonal $DyMnO_3$ exhibited intriguing spin-glass-like behavior. We demonstrated the possibility of enhancing the multiferroic properties of the lanthanide-based $R$MnO$_3$ by forming a metastable phase with a crystal structure different from that of their corresponding bulk materials.

This study was financially supported by the Korean Ministry of Science and Technology through the Creative Research Initiative program and by KOSEF through CSCMR. Work at Sungkyunkwan University was supported by the Korea Research Foundation (Grant No. 2005-C00153).

**Figure Captions**

Figure 1. $\theta$-$2\theta$ XRD scans of the DyMnO$_3$ films on (a) Pt(111)//Al$_2$O$_3$(0001) and (b) YSZ(111) substrates. The inset in Fig. 1(a) shows a rocking curve, which was recorded around the DyMnO$_3$ (0004) peak. Phi scans of the (11$\bar{2}$2) peaks for epitaxial hexagonal DyMnO$_3$ films on (c) Pt //Al$_2$O$_3$ (0001) and (d) YSZ(111) substrates.

Figure 2. Polarization *vs.* electric field hysteresis loops of the hexagonal DyMnO$_3$ films measured at (a) 25 and (b) 140 K. Dielectric constant *vs.* electric field curves are shown in the upper left insets. The d$^2P$/d$E^2$ *vs.* electric field plots are given in the lower right insets.

Figure 3. A contour map of the remnant polarization values as a function of temperature and maximum applied electric field. The dashed line shows a boundary between the ferroelectric and antiferroelectric-like regions determined as inflection points in the d$^2P$/d$E^2$ *vs.* electric field plots.

Figure 4. (a) Temperature-dependent magnetization curves measured with a 100 Oe magnetic field along the c-axis after zero-field cooling (ZFC) and field cooling (FC). (b) Temperature-dependent magnetization curves with a 100 Oe magnetic field applied along the substrate. As shown in the inset, the magnetization curve indicates two anomalies around 40 and 60 K, respectively. (c) Magnetic field-dependent magnetization curves at 4.2, 50, and 300 K. The inset shows an enlarged picture of the low-field data taken at 4.2 K with a clear hysteresis.

**References**

[1] J. Wang, J. B. Neaton, H. Zheng, V. Nagarajan, S. B. Ogale, B. Liu, D. Viehland, V. Vaithyanathan, D. G. Schlom, U. V. Waghmare, N. A. Spaldin, K. M. Rabe, M. Wuttig, and R. Ramesh, Science **299**, 1719 (2003).

[2] N. Hur, S. Park, P. A. Sharma, J. S. Ahn, S.Guha, and S. W. Cheong, Nature (London) **429**, 392 (2004).

[3] H. L. Yakel, W. D. Koehler, E. F. Bertaut, and F. Forrat, Acta Crystallogr. **16**, 957 (1963).

[4] T. Kimura, T. Goto, H. Shintani, K. Ishizaka, T. Arima, and Y. Tokura, Nature (London) **426**, 55 (2003).

[5] T. Goto, T. Kimura, G. Lawes, A. P. Ramirez, and Y. Tokura, Phys. Rev. Lett. **92**, 257201 (2004).

[6] T. Lottermoser, T. Lonkai, U. Amann, D. Hohlwein, J. Ihringer, and M. Fiebig, Nature **430**, 541 (2004).

[7] T. Katsufuji, S. Mori, M. Masaki, Y. Moritomo, N. Yamamoto, and H. Takagi, Phys. Rev. B 64, 104419 (2001).

[8] M. Fiebig, T. Lottermoser, D. Fröhlich, A. V. Goltsev, and R. V. Pisarev, Nature **419**, 818 (2002).

[9] B. B. V. Aken, T. T. M. Palstra, A. Filippetti, and N. A. Spaldin, Nature Mater. **3**, 164 (2004).

[10] A. A. Bosak, A. A. Kamenev, I. E. Graboy, S. V. Antonov, O. Y. Gorbenko, A. R. Kaul, C. Dubourdieu, J. P. Senateur, V. L. Svechnkikov, H. W. Zandbergen, and B. Holländer, Thin Solid Films **400**, 149 (2001).

[11] J.-H. Lee, P. Murugavel, H. J. Ryu, D. Lee, J. Y. Jo, J. W. Kim, H. J. Kim, K. H. Kim,

Y. Jo, M.-H. Jung, Y. W. Oh, Y.-W. Kim, J. G. Yoon, J.-S. Chung, and T. W. Noh, Adv. Mater. *To be published*.

[12] T. Lonkai, D. G. Tomuta, U. Amann, J. Ihringer, W. A. Hendrikx, D. M. Tobbens, J. A. Mydosh, Phys. Rev. B **69**, 134108 (2004).

[13] J. W. Park, J.-G. Park, G. S. Jeon, H. Y. Choi, C. H. Lee, W. Jo, r. Bewley, K. A. McEwen, and T. G. Perring, Phys. Rev. B **68**, 104426 (2003).

[14] B. Lorenz, F. Yen, M. M. Gospodinov, and C. W. Chu, Phys. Rev. B **71**, 014438 (2005).

[15] A. Munoz, J. A. Alonso, M. J. Martinez-Lope, M. T. Casais, J. L. Martinez, and M. T. Fernandez-Diaz, Phys. Rev. B **62**, 9498 (2000).

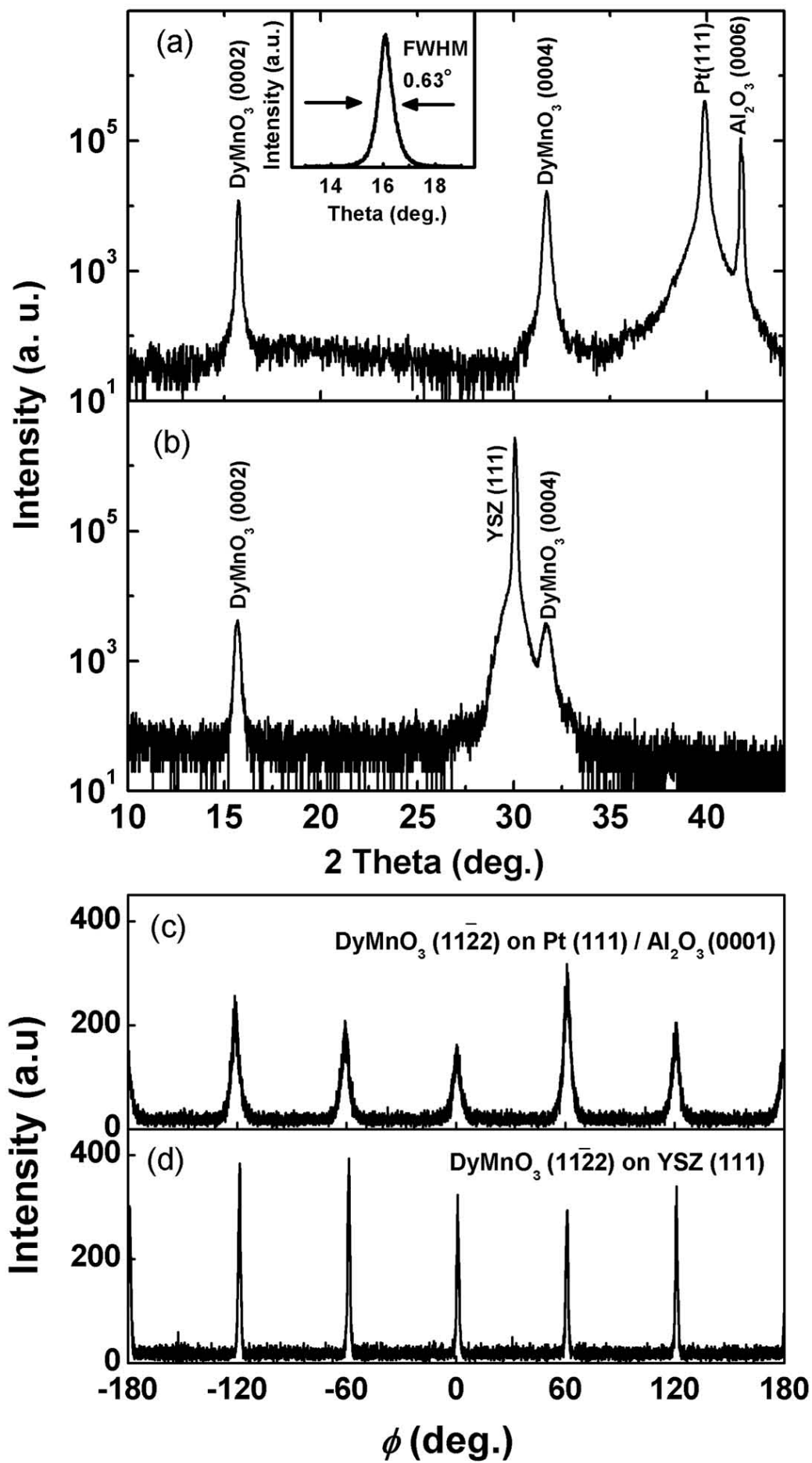

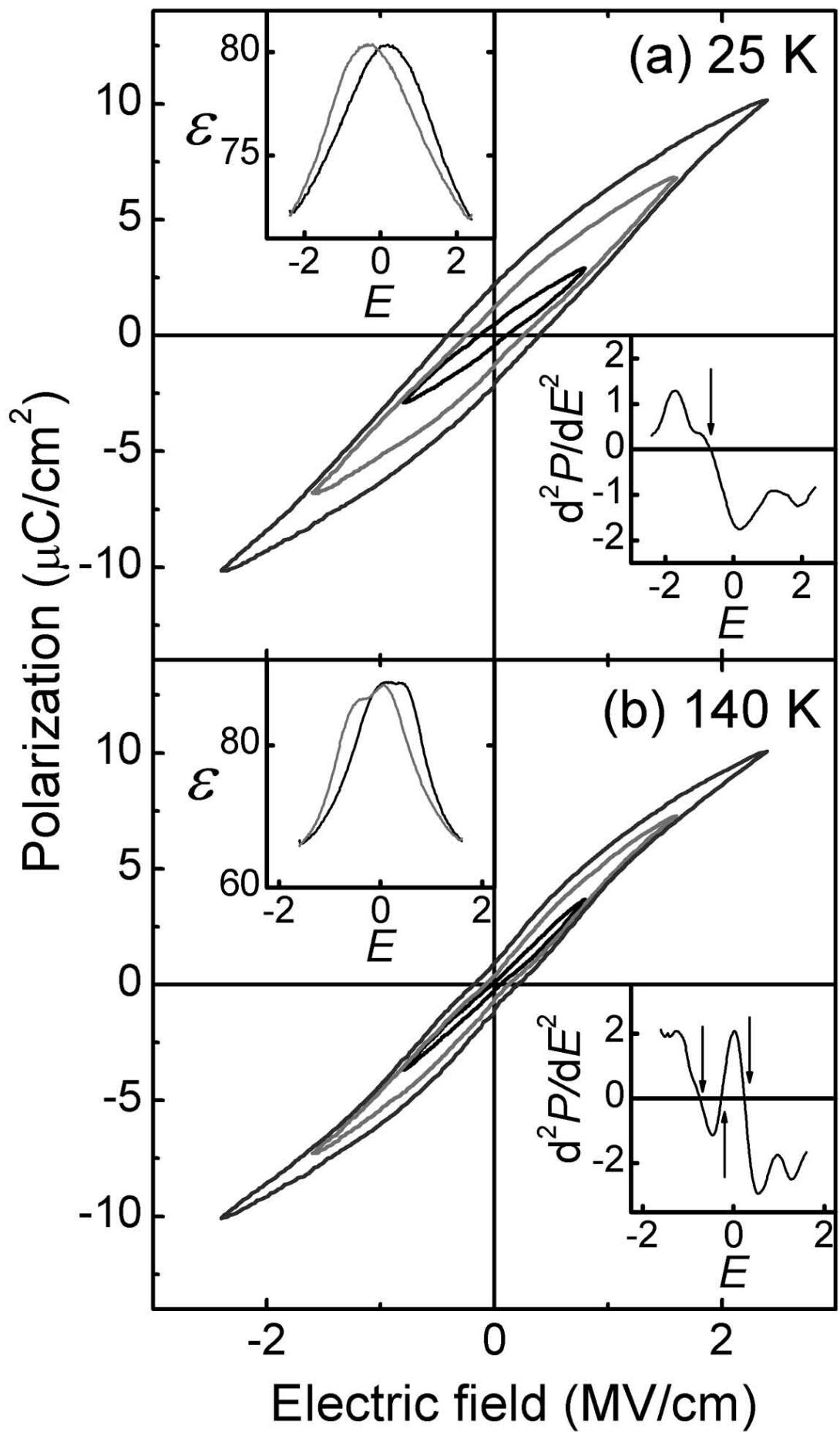

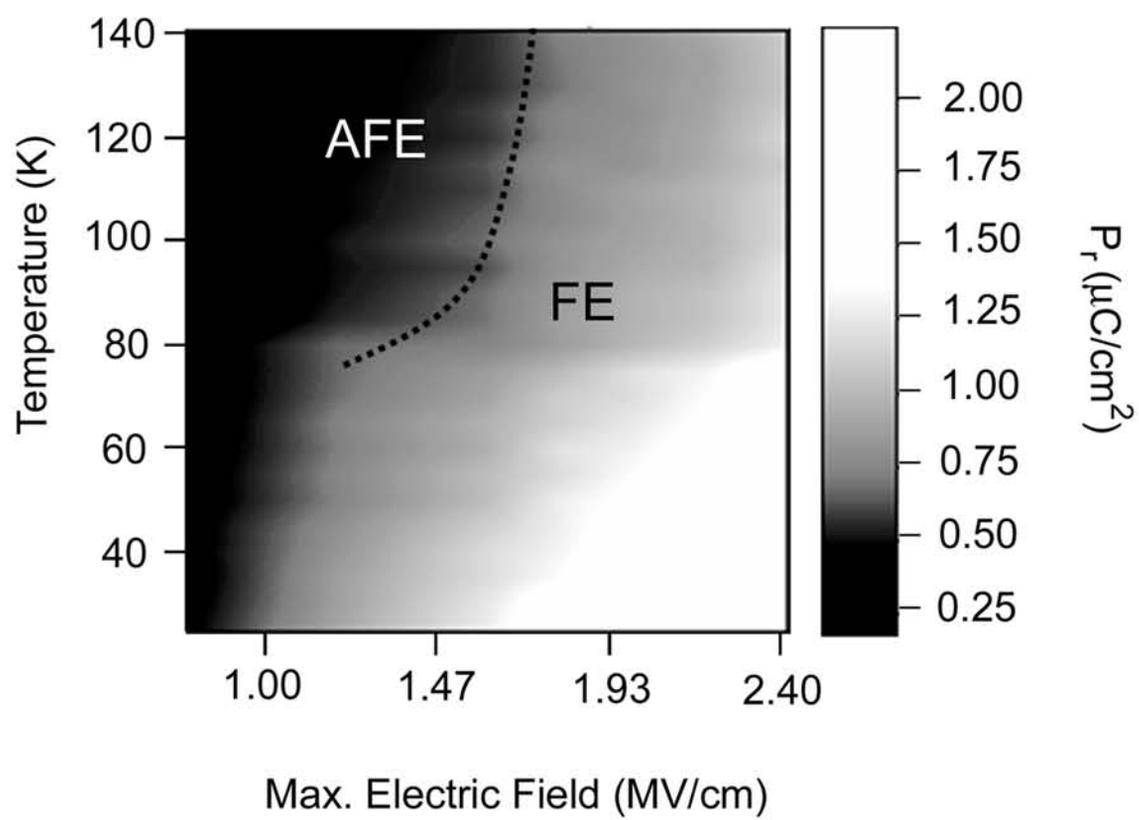

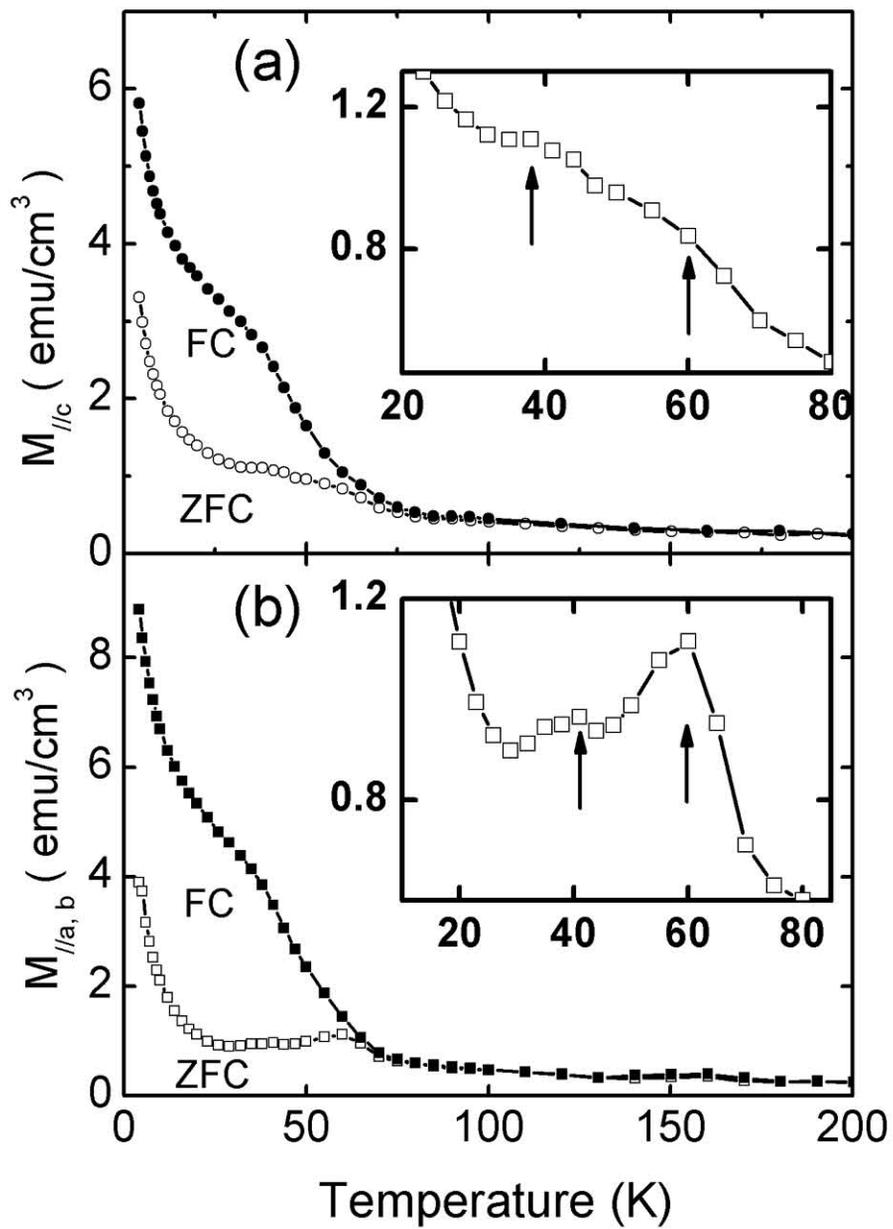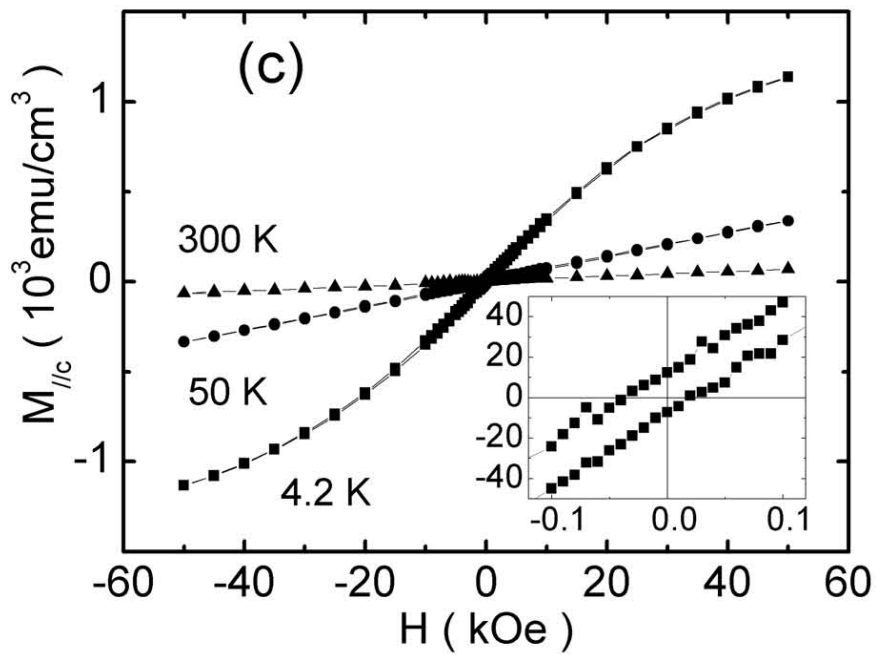